\documentclass{article}
\usepackage{spconf,amsmath,amssymb,amsfonts,graphicx,booktabs,textcomp,xcolor}
\usepackage{cite}
\usepackage[hidelinks]{hyperref}

\newcommand{\m}[2]{$#1_{\pm #2}$}
\newcommand{\ms}[3]{$#1^{#3}_{\pm #2}$}
\newcommand{\mB}[2]{$\mathbf{#1_{\pm #2}}$}
\newcommand{\msB}[3]{$\mathbf{#1^{#3}_{\pm #2}}$}
\newcommand{\mU}[2]{$\underline{#1_{\pm #2}}$}
\newcommand{\msU}[3]{$\underline{#1^{#3}_{\pm #2}}$}
\renewcommand{\arraystretch}{1.15}

\title{PHYSIOLOGY-INFORMED AUDIO QUALITY AND LISTENING EFFORT ESTIMATION}
\name{Niall McGuire$^{1}$\thanks{Work done during an internship at Microsoft in WA, USA.} \qquad Dimitra Emmanouilidou$^{2}$}
\address{$^{1}$University of Strathclyde, Glasgow, UK \\
  $^{2}$Microsoft Research, Redmond, WA, USA}

\begin{document}
\ninept
\maketitle

\begin{abstract}
Audio systems aim to maximise the listener's perceived quality, traditionally evaluated through mean opinion scoring (MOS), a subjective process that is slow, costly, and collected only after playback. Machine learning now predicts MOS directly from the audio, but in generalising across participants it tends to discard the subject variance that defines individual perception. Neurophysiological measures may offer a complementary signal taken directly from the listener, objective, implicit, and continuous, though prior work has relied on laboratory-grade EEG and fNIRS. We ask how practical consumer-grade devices compare, and whether they might recover population-level audio-quality trends and, when calibrated per listener, track that listener's effort. Recording 32-channel laboratory EEG alongside wearable EEG, fNIRS, and eye tracking from thirty listeners, our results suggest commercial wearable and optical sensing can decode audio quality at statistically significant levels and recover listening effort per listener. These findings may point towards platforms tuned not only to perceived quality but to the effort a clip demands, read from the listener's own response as they listen.
\end{abstract}


\section{Introduction}

Audio systems aim to maximise perceived quality, traditionally evaluated through mean opinion scoring (MOS), where listeners rate clips and scores are averaged into a single value. Whilst long the industry standard \cite{itu_p800_1996}, MOS has clear limits. A listener's rating depends on personal preference, prior experience, and a shifting interpretation of the scale, producing substantial variability across listeners and sessions \cite{streijl2016mos, maki2018tensor}. Due to this variance, it requires large, costly panels, and as ratings are gathered only after playback, MOS cannot support automated or adaptive evaluation~\cite{li2026brain}.
Machine learning methods address these limitations by predicting MOS directly from audio, providing fast, reproducible, human-free labelling. Reference-based metrics such as PESQ \cite{rix2001pesq} and SDR \cite{vincent2006sdr} compare degraded audio against a clean reference unavailable during live playback, while reference-free models such as DNSMOS for speech \cite{reddy2021dnsmos} and Fr\'echet Audio Distance for music and generated audio \cite{kilgour2018frechet} operate on the degraded signal alone. However, these methods still predict a single population-average score, discarding the subject-specific variation that defines individual perception \cite{maki2018tensor, li2026brain}.

Neurophysiological measures provide a complementary source of information by capturing the listener's ongoing perceptual and cognitive response to audio stimuli \cite{engelke2017qoe, gupta2015physyqx}. The most practical for everyday use are electroencephalography (EEG) and functional near-infrared spectroscopy (fNIRS), compact and movement-tolerant in ways functional magnetic resonance imaging (fMRI) is not. EEG resolves activity at the millisecond scale, matching the temporal structure of perceived quality, while the slower haemodynamic response of fNIRS may track effort or demand~\cite{maki2018tensor, vaisberg2024fnirs}. Prior work has shown that speech degradations alter EEG responses and that physiological measures can be used to estimate perceived speech quality directly \cite{antons2012eeg, arndt2014interspeech}. Recorded as the participant listens, such signals are objective in not requiring the listener's conscious judgement, implicit and continuous, indexing perceptual and cognitive state without interrupting for a score rating~\cite{maki2018tensor,li2026brain,song2025eeg}. 

Physiology also exposes a dimension a single rating collapses into one value: the cognitive effort the audio costs the listener. Two clips given the same quality rating can demand very different processing, and that hidden effort accumulates into fatigue and disengagement over time \cite{winn2015spectral, pichorafuller2016fuel}. Pupil dilation grows with spectral degradation even on trials that reach the same correct answer, a cost a quality rating does not register \cite{winn2015spectral, zekveld2018pupil}, and effortful listening to continuous speech raises pupil dilation while lowering parietal alpha power \cite{seifiala2020alpha}. Additionally, pupillometry and EEG could each capture effort but perhaps different components of it \cite{miles2017coregistration}. And for a real-time adaptive audio system, measuring quality and effort can help personalise experiences to the listener, both in terms of quality perception and effort to process~\cite{vaisberg2024fnirs, simeone2025alpha}.

This line of work, however, rests on laboratory-grade equipment. From the early Berlin studies to recent EEG-MOS models, high-density wet-electrode systems tether the listener to the laboratory \cite{antons2012eeg, maki2018tensor, li2026brain, song2025eeg}, with the strongest results reported from subject-specific models. Whether perceived quality survives on low-channel, dry-electrode consumer hardware, and whether quality and effort can be decoded simultaneously, still remains unclear. Our contribution is therefore narrow and feasibility-focused: a simultaneous comparison of laboratory and consumer sensing (wearable EEG, fNIRS, and eye tracking) on the same listeners, and a joint treatment of audio quality and listening effort. We do not claim physiology outperforms signal-based metrics, which the evidence does not support \cite{maki2018tensor}. We address three questions:
\begin{itemize}
\item \textbf{RQ1:} to what extent can neurophysiological signals recover audio quality?
\item \textbf{RQ2:} does degraded audio increase listening effort, and can that effort be decoded from physiology?
\item \textbf{RQ3:} how does decoding performance with consumer-grade wearable EEG, fNIRS, and eye tracking compare to laboratory-grade EEG?
\end{itemize}
To answer them we present listeners with graded-quality audio while recording 32-channel laboratory EEG, wearable EEG, fNIRS, and eye tracking, and relate the resulting features to the audio's objective quality labels and to each listener's effort ratings, evaluated across unseen and calibrated listeners.

\section{Methods}
\subsection{Participants, Apparatus, and Stimuli}

Thirty listeners took part, of whom 21 completed the pre-session questionnaire (16 male, 5 female, aged 33.3 $\pm$ 11.6 years, 9 native English speakers, and 7 with prior listening-test experience). Recruited participants reported no history of neurological disorder or concussion within the preceding six months, and each received a \$10 gift card. The study held institutional ethics approval and all participants gave written informed consent. All thirty were considered for every task, with the eligible sample reported per task since a listener contributes only when a valid label and the relevant sensor stream are both present.

We recorded four physiological streams from three devices, two of them commercially available wearables. The laboratory reference was a 32-electrode Brain Products LiveAmp at 256 Hz in the standard 10/20 montage. Wearable EEG and fNIRS came from a Muse Athena headband, an off-the-shelf consumer device, providing the four scalp channels TP9, AF7, AF8, and TP10 at 256 Hz and fNIRS at 64 Hz from four source-detector pairs at 730 and 850 nm. Eye tracking used a Pupil Labs Neon, a commercially available wearable eye tracker, recording binocular gaze and pupillometry at 200 Hz. All streams were synchronised through the Lab Streaming Layer and stored as XDF, with stimulus onsets and offsets marked from the presentation software. Audio was played over M-Audio Studiophile BX8 near-field monitors, with no spatial manipulation beyond the reverberation injected into a controlled subset of degraded clips, and rating prompts shown on a 16-inch laptop, in a quiet, windowless room with constant ambient lighting. Before playback, every clip was loudness-normalised using A-weighting,
 and playback gain held fixed across sessions, so the level did not covary with quality.

Stimuli were drawn from established MOS-annotated corpora, NISQA for degraded speech \cite{mittag2021nisqa} and ODAQ for music \cite{torcoli2024odaq}, spanning codec, additive-noise, reverberation, and packet-loss conditions. The graded set was speech only, sampled to populate all five objective MOS categories with the corpus MOS as the graded quality target. Separate extreme blocks presented clean and degraded clips whose clean or degraded label defines the binary quality target, with the degraded clips balanced at five per artefact type within each content family. In total 144 clips were presented, 102 speech and 42 music, each about 10 $\pm$ 2 s long.

\begin{table*}[t]
\centering
\caption{Binary tasks (AQ-2, EF-2), balanced accuracy against chance $0.5$, higher better, mean$_{\pm\mathrm{SD}}$ over folds. Best per column \textbf{bold}, second \underline{underlined}. LOSO tests an unseen listener, WS a calibrated one. Full uses all of a modality's trials, matched only those common to all four. $^{\ast}/^{\ast\ast}/^{\ast\ast\ast}$ mark Holm $p{<}.05/.01/.001$ vs.\ chance, $\downarrow$ below.\\}
\label{tab:bin}
\footnotesize
\setlength{\tabcolsep}{2pt}
\begin{tabular}{l rr cc | rr cc @{\hskip 10pt} rr cc | rr cc}
\toprule
& \multicolumn{8}{c}{AQ-2} & \multicolumn{8}{c}{EF-2} \\
\cmidrule(lr){2-9}\cmidrule(lr){10-17}
& \multicolumn{4}{c|}{Full} & \multicolumn{4}{c}{Matched} & \multicolumn{4}{c|}{Full} & \multicolumn{4}{c}{Matched} \\
\cmidrule(lr){2-5}\cmidrule(lr){6-9}\cmidrule(lr){10-13}\cmidrule(lr){14-17}
Modality & $N$ & Trials & LOSO & WS & $N$ & Trials & LOSO & WS & $N$ & Trials & LOSO & WS & $N$ & Trials & LOSO & WS \\
\midrule
Muse EEG & 25 & 1023 & \msU{0.64}{.12}{***} & \msB{0.67}{.05}{***} & 8 & 122 & \mU{0.55}{.23} & \m{0.48}{.11} & 28 & 1613 & \m{0.52}{.13} & \msU{0.61}{.04}{***} & 11 & 236 & \mU{0.52}{.19} & \mB{0.59}{.05} \\
fNIRS    & 17 & 610  & \ms{0.60}{.09}{***} & \ms{0.58}{.04}{***} & 8 & 122 & \msB{0.65}{.11}{***} & \mU{0.53}{.08} & 21 & 1235 & \ms{0.52}{.04}{*} & \ms{0.58}{.02}{**} & 11 & 236 & \m{0.48}{.18} & \mU{0.52}{.05} \\
Eye      & 25 & 1164 & \msB{0.65}{.17}{***} & \msU{0.66}{.04}{***} & 8 & 122 & \m{0.44}{.09} & \mB{0.54}{.06} & 29 & 1891 & \mB{0.57}{.17} & \msB{0.67}{.02}{***} & 11 & 236 & \mU{0.52}{.07} & \m{0.50}{.04} \\
Lab EEG  & 12 & 218  & \m{0.51}{.09} & \ms{0.37}{.08}{**\downarrow} & 8 & 122 & \m{0.50}{.07} & \m{0.47}{.07} & 18 & 381 & \mU{0.56}{.23} & \m{0.58}{.04} & 11 & 236 & \mB{0.56}{.17} & \mB{0.59}{.12} \\
\bottomrule
\end{tabular}
\end{table*}

\subsection{Procedure}

Each session opened with anchor clips, one per artefact type and MOS level, showing the range of degradations before rating began, a step added after pilot ratings drifted toward extremes never actually encountered. Trials then fell into two families. On graded trials the listener rated clip quality on a five-position scale and listening effort on a separate rating scale, and on extreme trials they made a categorical high or poor quality judgement alongside an ordinal effort rating. Throughout, a concurrent visual task tracking zero, one, or three moving targets sustained engagement, its load randomised independently of audio quality and its responses recorded but not modelled here. Across both families the effort ratings supply the effort targets, whereas the quality targets come from the fixed corpus MOS and the clean or degraded label, not the listener's own rating, a choice quantified in Table~\ref{tab:target}. Artefact type and MOS level were each presented in randomised order.

\subsection{Physiological Processing and Features}

All streams were epoched from a two-second prestimulus baseline and baseline-corrected, to the audio offset for EEG, gaze, and pupillometry and to a fixed post-offset interval for fNIRS. EEG was line-notched, band-pass filtered from 1 to 45 Hz, common-average referenced, and cleaned of ocular and high-kurtosis components by FastICA \cite{hyvarinen2000ica}, with laboratory-EEG trials after cap removal discarded. fNIRS intensities were converted to optical density \cite{delpy1988optical}, motion-corrected \cite{scholkmann2010motion}, band-pass filtered from 0.01 to 0.2 Hz, and summarised over a fixed-latency window aligned to the canonical auditory haemodynamic response, which peaks roughly five to seven seconds after onset \cite{scholkmann2014review}. The Muse's four forehead optodes sit over prefrontal cortex rather than auditory cortex, so its fNIRS indexes frontal engagement and listening effort more than auditory quality processing directly, and carrying no short-separation reference, systemic haemodynamics may mix with the cortical signal. The band-pass and baseline attenuate but do not fully remove it, so we read fNIRS as a feasible optical channel rather than a clean cortical measure. Pupil traces were validated against the device worn flag, gap-interpolated, and expressed as proportional change from the prestimulus median. Per EEG channel we took absolute and relative delta-to-gamma band power, total power, spectral entropy, peak frequency, and the upper triangle of a shrinkage channel covariance \cite{chen2010shrinkage}. Each fNIRS channel contributed amplitude, slope, area under the curve, time to peak, and early, middle, and late means, and the eye tracker gaze velocity, dispersion, and path length alongside pupil dilation amplitude, slope, time to peak, and binocular agreement. 

\subsection{Prediction Tasks and Evaluation}

We evaluate four tasks, two for audio quality (AQ) and two for listening effort (EF), each as a binary contrast (-2) and a five-point target (-5). For quality, AQ-2 separates extreme clean from extreme degraded clips by their objective label and AQ-5 regresses the five-point corpus MOS. For effort, taken from the listener's rating, EF-2 separates low (1--2) from high (4--5) ratings and EF-5 regresses the five-point rating. The binary tasks are fit by minimising binary cross-entropy and the regression tasks by minimising mean squared error, each against the targets above.

We score binary tasks by balanced accuracy against a chance level of 0.5, and continuous tasks by the Pearson correlation against a null of zero and by the root mean squared error (RMSE) in rating units, reporting mean and standard deviation across outer folds and testing significance for balanced accuracy and correlation only. Within each outer training fold, model family and hyperparameters were chosen by three-fold participant-grouped cross-validation over SVM, elastic-net, XGBoost, MLP, and CatBoost, each searched over standard ranges, adding PLS for regression and an OAS-covariance \cite{chen2010shrinkage} Riemannian tangent-space model \cite{barachant2012riemannian} for EEG, with the outer test fold never informing selection. Two outer regimes answer different questions. Leave-one-subject-out (LOSO) holds out one listener entirely, measuring generalisation to unseen listeners, whereas a pooled five-fold within-subject (WS) regime holds out a fifth of each listener's trials, so the score reflects subject-dependent prediction. Its correlation is pooled across listeners within a fold rather than averaged per listener, so a stable between-listener offset can inflate it. Both regimes reuse the fixed clip set, so they establish cross-listener transfer, not generalisation to unseen audio: because the clips recur, a decoder may partly exploit stimulus-evoked responses rather than perceived quality, which a leave-one-clip-out regime, not attempted here, would isolate. Relatedly, the corpus MOS we regress is a population label, so decoding it need not mean decoding a given listener's perceived quality, as the within-listener analyses below show. Each modality is reported under full support (all its trials) and matched support (trials common to all four modalities), which equalises the cohort at the cost of a smaller sample. Significance used two-sided participant-level bootstrap tests with 50,000 resamples, Holm-corrected within each task across its eight tests. A marker reports whether a model differs from its own null, not whether modalities or support conditions differ, which we do not test. This design departs from prior EEG-quality decoding studies in three respects: hyperparameters are chosen without outer-fold leakage, trial and listener counts are equalised via the matched support condition rather than left to vary freely with modality, and a below-chance result (laboratory EEG, see next section) is reported rather than omitted. To our knowledge, prior work has not controlled for all three jointly.

\begin{table*}[t]
\centering
\caption{Regression tasks (AQ-5, EF-5), Pearson $r$ against null $0$, higher better, mean$_{\pm\mathrm{SD}}$ over folds. Best per column \textbf{bold}, second \underline{underlined}. Support, regimes, and markers as in Table~\ref{tab:bin}. Full-support RMSE, untested, spans 1.18--1.30 MOS units (AQ-5) and 0.83--0.96 rating units (EF-5).\\}
\label{tab:reg}
\footnotesize
\setlength{\tabcolsep}{2pt}
\begin{tabular}{l rr cc | rr cc @{\hskip 10pt} rr cc | rr cc}
\toprule
& \multicolumn{8}{c}{AQ-5} & \multicolumn{8}{c}{EF-5} \\
\cmidrule(lr){2-9}\cmidrule(lr){10-17}
& \multicolumn{4}{c|}{Full} & \multicolumn{4}{c}{Matched} & \multicolumn{4}{c|}{Full} & \multicolumn{4}{c}{Matched} \\
\cmidrule(lr){2-5}\cmidrule(lr){6-9}\cmidrule(lr){10-13}\cmidrule(lr){14-17}
Modality & $N$ & Trials & LOSO & WS & $N$ & Trials & LOSO & WS & $N$ & Trials & LOSO & WS & $N$ & Trials & LOSO & WS \\
\midrule
Muse EEG & 29 & 3183 & \msB{0.38}{.14}{***} & \msB{0.43}{.03}{***} & 12 & 573 & \ms{0.07}{.10}{*} & \m{0.23}{.04} & 28 & 3193 & \m{0.09}{.23} & \msU{0.23}{.05}{***} & 11 & 526 & \mB{0.02}{.19} & \mB{0.17}{.08} \\
fNIRS    & 22 & 2417 & \msU{0.36}{.11}{***} & \msU{0.41}{.05}{***} & 12 & 573 & \msB{0.19}{.15}{***} & \msB{0.34}{.03}{***} & 21 & 2400 & \m{0.02}{.11} & \ms{0.13}{.06}{***} & 11 & 526 & \m{-0.13}{.31} & \m{0.07}{.05} \\
Eye      & 30 & 3750 & \ms{0.20}{.29}{**} & \msU{0.41}{.04}{***} & 12 & 573 & \msU{0.10}{.15}{*} & \mU{0.29}{.04} & 29 & 3761 & \mB{0.14}{.37} & \msB{0.37}{.03}{**} & 11 & 526 & \mU{-0.10}{.19} & \mU{0.11}{.07} \\
Lab EEG  & 19 & 867  & \m{0.05}{.33} & \ms{0.15}{.05}{*} & 12 & 573 & \m{-0.06}{.15} & \m{0.13}{.22} & 18 & 816 & \mU{0.12}{.44} & \m{0.13}{.09} & 11 & 526 & \m{-0.18}{.35} & \m{0.08}{.14} \\
\bottomrule
\end{tabular}
\end{table*}

\section{Results and Discussion}

\begin{figure}[t]
\centering
\includegraphics[width=0.49\linewidth]{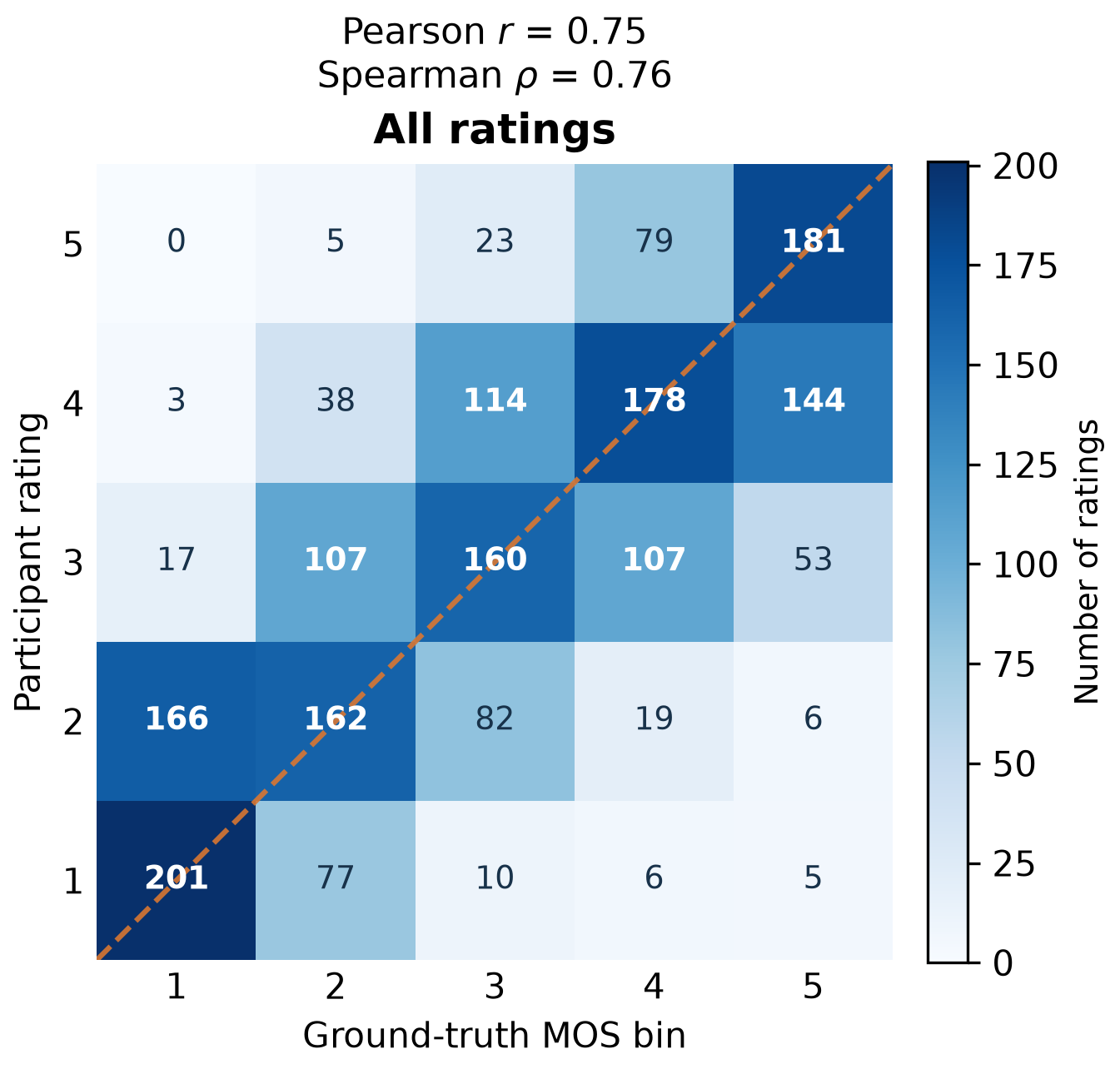}\hfill
\includegraphics[width=0.49\linewidth]{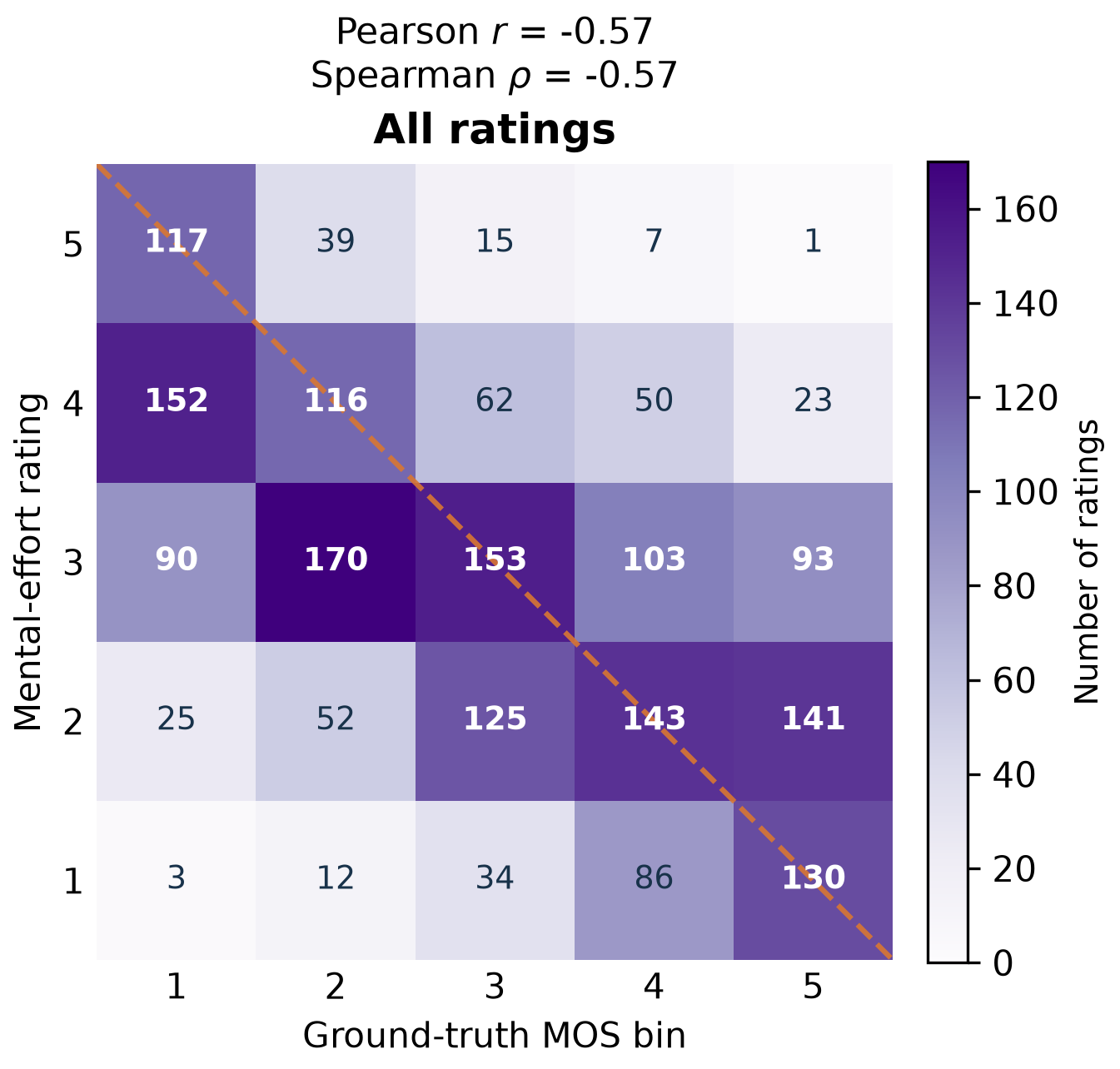}
\caption{Behavioural validation against objective MOS. Left: subjective quality tracks MOS ($r{=}0.75$). Right: mental effort rises as quality falls ($r{=}{-}0.57$). Cells are rating counts, dashed line marks agreement.}
\label{fig:behaviour}
\end{figure}

\noindent\textbf{Behavioural validation.}
We first check that listeners perceived the intended quality and that it carried an effort cost. Their subjective quality ratings track the corpus MOS closely (Pearson $r$ 0.75, Spearman $\rho$ 0.76), so each clip appears heard at its labelled quality, supporting the corpus MOS as the modelling target. Mental-effort ratings move the opposite way, rising as quality falls ($r$ $-0.57$), so degraded clips also seem to have demanded more effort even though the two were collected as separate judgements (Figure~\ref{fig:behaviour}).

\noindent\textbf{Recovering audio quality (RQ1).}
Objective quality appears recoverable from physiology (Tables~\ref{tab:bin} and~\ref{tab:reg}). On the binary contrast, full-support wearable EEG, fNIRS, and eye tracking exceed chance for unseen listeners at balanced accuracy 0.60 to 0.65, and again when calibrated to the listener. The graded five-point target seems harder but still tractable: correlations with the corpus MOS are positive and significant, strongest for wearable EEG and fNIRS under LOSO ($r$ 0.38 and 0.36, RMSE near 1.18 to 1.20 MOS points) and rising to 0.41 to 0.43 under subject-dependent prediction. Because the clip set recurs across folds, these figures reflect recovery of the corpus label under cross-listener transfer. The 32-channel laboratory EEG appears the exception throughout, neither exceeding chance on the binary task nor correlating with MOS for unseen listeners, a pattern we return to below.

\noindent\textbf{From laboratory to wearable sensing (RQ3).}
The wearable and optical streams stay above chance in both regimes, whereas the 32-channel stream does not exceed chance on the binary contrast for unseen listeners. We cannot claim that the wearables outperform laboratory EEG, because the comparison is additionally confounded by the fewer trials of the laboratory stream, its different channel count, a mid-study protocol change, and a wet-electrode cap harder to hold at stable impedance than the dry wearables. The one below-chance cell, laboratory EEG at 0.37, likewise need not be a result, as train and test distribution shift, contact instability, or a pipeline artefact could each produce it. The safer reading is that laboratory recording carries real practical costs in this setting, while the wearable and optical streams still recover quality with adequate promise. The matched condition equalises trials across all four modalities, and there fNIRS is the only modality to exceed chance (0.65) and to correlate with MOS ($r$ 0.19) for unseen listeners, so we treat fNIRS as the most consistent modality once support is equalised.

\noindent\textbf{Decoding listening effort (RQ2).}
The behavioural analysis already suggested that degraded audio raises rated effort. Tables~\ref{tab:bin} and~\ref{tab:reg} then ask whether that effort is decodable, and point to recovery mainly under subject-dependent prediction. Wearable EEG, fNIRS, and eye tracking then appear to separate low from high effort and to track the graded rating, eye tracking most strongly ($r$ 0.37, RMSE 0.89). Across unseen listeners the signal does not yet appear to generalise: no modality decodes the five-point rating after correction, and only fNIRS reaches significance on the binary contrast, with a small effect at 0.52. Effort therefore seems to leave a physiological trace a model can read once it has seen the listener, but not one that transfers to new listeners here. One caveat applies even within listener: rated effort correlates negatively with quality ($r$ $-0.57$), so an effort decoder may partly read the quality contrast. The concurrent visual load is a second possible confound for the gaze and pupil streams, though its level was randomised independently of quality and the added multi-target load appeared on only a minority of trials, so we expect little conflation of visual with listening effort. Both are left to follow-up.

\begin{table}[t]
\centering
\renewcommand{\arraystretch}{1.0}
\caption{Binary quality (AQ-2) by content, speech (Sp) vs.\ music (Mu). Participant-mean balanced accuracy against chance $0.5$, mean$_{\pm\mathrm{SD}}$, best per column \textbf{bold}, second \underline{underlined}. $P$ is the speech/music listener count. Descriptive, single seed.\\}
\label{tab:content}
\footnotesize
\setlength{\tabcolsep}{1pt}
\begin{tabular}{lcccccc}
\toprule
& & & \multicolumn{2}{c}{LOSO} & \multicolumn{2}{c}{WS} \\
\cmidrule(lr){4-5}\cmidrule(lr){6-7}
Modality & Trials & $P_\mathrm{sp}/P_\mathrm{mu}$ & Sp & Mu & Sp & Mu \\
\midrule
Muse EEG & 334 & 11/23 & \mB{0.83}{.12} & \mU{0.57}{.14} & \mB{0.83}{.09} & \m{0.50}{.21} \\
fNIRS    & 96  & 3/16  & \m{0.75}{.03} & \m{0.48}{.20} & \mU{0.73}{.05} & \mU{0.53}{.22} \\
Eye      & 352 & 11/25 & \mB{0.83}{.16} & \mB{0.59}{.21} & \mB{0.83}{.15} & \mB{0.62}{.19} \\
Lab EEG  & 34  & 4/9   & \mU{0.78}{.06} & \m{0.38}{.24} & \m{0.69}{.10} & \m{0.52}{.27} \\
\bottomrule
\end{tabular}
\end{table}

\begin{table}[t]
\centering
\renewcommand{\arraystretch}{1.0}
\caption{Quality supervised by the listener's rating (Subj) or the corpus MOS (Obj), both scored against corpus MOS. RMSE in MOS units (lower better) and Pearson $r$ (higher better), mean$_{\pm\mathrm{SD}}$. The Obj arm is the deployed model of Table~\ref{tab:reg}, so the gap bounds the benefit of objective supervision rather than isolating it. Descriptive, single seed.\\}
\label{tab:target}
\footnotesize
\setlength{\tabcolsep}{2.5pt}
\begin{tabular}{lcccccc}
\toprule
& & & \multicolumn{2}{c}{RMSE} & \multicolumn{2}{c}{$r$} \\
\cmidrule(lr){4-5}\cmidrule(lr){6-7}
Modality & $N$ & Trials & Subj & Obj & Subj & Obj \\
\midrule
\multicolumn{7}{l}{\textit{LOSO}} \\
Muse EEG & 29 & 3183 & \m{1.38}{.05} & \m{1.20}{.09} & \m{0.19}{.13} & \m{0.38}{.14} \\
fNIRS    & 22 & 2417 & \m{1.38}{.04} & \m{1.18}{.07} & \m{0.18}{.11} & \m{0.36}{.11} \\
Eye      & 30 & 3750 & \m{1.35}{.16} & \m{1.24}{.16} & \m{0.19}{.30} & \m{0.20}{.29} \\
Lab EEG  & 19 & 867  & \m{1.47}{.11} & \m{1.30}{.16} & \m{0.01}{.47} & \m{0.05}{.33} \\
\addlinespace
\multicolumn{7}{l}{\textit{Within-subject}} \\
Muse EEG & 29 & 3183 & \m{1.39}{.02} & \m{1.18}{.00} & \m{0.16}{.06} & \m{0.43}{.03} \\
fNIRS    & 22 & 2417 & \m{1.38}{.00} & \m{1.18}{.03} & \m{0.19}{.01} & \m{0.41}{.05} \\
Eye      & 30 & 3750 & \m{1.36}{.01} & \m{1.19}{.05} & \m{0.16}{.08} & \m{0.41}{.04} \\
Lab EEG  & 19 & 867  & \m{1.46}{.04} & \m{1.26}{.05} & \m{0.01}{.06} & \m{0.15}{.05} \\
\bottomrule
\end{tabular}
\end{table}

\noindent\textbf{Supplementary analyses.}
Two descriptive checks probe the quality tasks. Within-subject figures are pooled across listeners within a fold, as noted above, so they reward a shared population axis. First, on the supervising label (Table~\ref{tab:target}), training on the objective corpus MOS rather than the listener's rating lowered pooled RMSE against objective MOS in every case and raised the pooled correlation, clearest under subject-dependent prediction, where $r$ rose from around 0.17 to between 0.41 and 0.43 for wearable EEG, fNIRS, and eye tracking. A fuller model search sat behind the objective arm, so we read this as an upper bound rather than a controlled contrast. Second, this pooled gain appears to be shrinkage rather than denoising. Objective-trained predictions vary far less than the raw ratings among clips of equal objective MOS (within-group dispersion falling 55 to 94 percent), yet scored within listener they worsen, rank correlation falling from around 0.79 to between 0.12 and 0.33. Read together, objective supervision may recover the population axis the pooled metrics reward by regressing each clip towards its group mean, which within-listener ranking penalises. We therefore supervise the deployed models on the objective label, leaving a single listener's ordering as an open problem.

\noindent\textbf{Speech versus music.}
Splitting the extreme binary contrast by content suggests the quality signal is carried mostly by speech (Table~\ref{tab:content}). This descriptive, single-seed run matches clean and degraded trials within each listener-content block, so the families differ only in their listeners. Speech then exceeds music in every modality, at participant-mean balanced accuracy 0.69 to 0.83 against 0.38 to 0.62, several music cells at or below chance, with the pooled accuracies of Table~\ref{tab:bin} in between. Equalising trials left some families with very few listeners (three for speech fNIRS, four for laboratory EEG), so this is a descriptive pattern rather than a controlled effect.

\section{Conclusion}

Physiological sensing offers tentative support on all three questions. Binary quality can be decoded above chance from wearable EEG, fNIRS, and eye tracking across unseen listeners, graded MOS at lower but significant correlations, while the laboratory EEG shows the practical costs of high-density recording and fNIRS is most consistent once support is equalised. Effort appears decodable per listener, eye tracking and fNIRS most strongly. These signals recover the corpus-level quality axis, and, once calibrated to a listener, that listener's effort ratings. Whether they can also recover a single listener's own quality ordering, rather than only this shared population axis, remains open.
These are feasibility results, not a claim to replace signal-based metrics: correlations are modest, effort does not yet transfer to unseen listeners, and the between-modality comparisons rest on uneven cohorts. Scoring the speech clips with a predictor such as DNSMOS to bound the stimulus ceiling, and a leave-one-clip-out test of generalisation to unseen audio, are natural next steps. Even so, the results hint at systems tuned to perceived quality and processing effort alike.
\bibliographystyle{IEEEbib}
\bibliography{references}

@misc{itu_p800_1996,
  author       = {{International Telecommunication Union}},
  title        = {Methods for Subjective Determination of Transmission Quality},
  howpublished = {ITU-T Recommendation P.800},
  year         = {1996}
}

@article{streijl2016mos,
  author  = {Streijl, Robert C. and Winkler, Stefan and Hands, David S.},
  title   = {Mean opinion score ({MOS}) revisited: Methods and applications, limitations and alternatives},
  journal = {Multimedia Systems},
  volume  = {22}, number = {2}, pages = {213--227}, year = {2016}
}

@article{maki2018tensor,
  author  = {Maki, Hayato and Sakti, Sakriani and Tanaka, Hiroki and Nakamura, Satoshi},
  title   = {Quality prediction of synthesized speech based on tensor structured {EEG} signals},
  journal = {PLOS ONE},
  volume  = {13}, number = {6}, pages = {e0193521}, year = {2018}
}

@inproceedings{li2026brain,
  author    = {Li, Mingyu and Hu, Shuzhan and Liu, Yang and Liu, Danjing and Zhong, Wei and Ye, Long},
  title     = {Brain-inspired audio quality assessment based on audio-{EEG} feature fusion},
  booktitle = {Proc. Int. Forum Digital TV and Wireless Multimedia Commun. (IFTC 2025)},
  series    = {CCIS}, volume = {2902}, pages = {281--293}, publisher = {Springer}, year = {2026}
}

@inproceedings{rix2001pesq,
  author    = {Rix, Antony W. and Beerends, John G. and Hollier, Michael P. and Hekstra, Andries P.},
  title     = {Perceptual evaluation of speech quality ({PESQ})},
  booktitle = {Proc. IEEE ICASSP}, volume = {2}, pages = {749--752}, year = {2001}
}

@inproceedings{reddy2021dnsmos,
  author    = {Reddy, Chandan K. A. and Gopal, Vishak and Cutler, Ross},
  title     = {{DNSMOS}: A non-intrusive perceptual objective speech quality metric to evaluate noise suppressors},
  booktitle = {Proc. IEEE ICASSP}, pages = {6493--6497}, year = {2021}
}

@article{engelke2017qoe,
  author  = {Engelke, Ulrich and Darcy, Daniel P. and Mulliken, Grant H. and Bosse, Sebastian and Martini, Maria G. and Arndt, Sebastian and Antons, Jan-Niklas and Chan, Kit Yan and Ramzan, Naeem and Brunnstr{\"o}m, Kjell},
  title   = {Psychophysiology-based {QoE} assessment: A survey},
  journal = {IEEE J. Sel. Topics Signal Process.},
  volume  = {11}, number = {1}, pages = {6--21}, year = {2017}
}

@inproceedings{gupta2015physyqx,
  author    = {Gupta, Rishabh and Banville, Hubert J. and Falk, Tiago H.},
  title     = {{PhySyQX}: A database for physiological evaluation of synthesised speech quality-of-experience},
  booktitle = {Proc. IEEE WASPAA}, pages = {1--5}, year = {2015}
}

@inproceedings{song2025eeg,
  author    = {Song, Qiwei and Xiang, Tao and Ge, Ning and Meng, Hao and Wei, Junyi},
  title     = {An electroencephalography-based method for audio perception assessment},
  booktitle = {Proc. Int. Conf. Ubiquitous Communication (Ucom)}, pages = {315--319}, year = {2025}
}

@article{vaisberg2024fnirs,
  author  = {Vaisberg, Jonathan M. and Gilmore, Sean and Qian, Jinyu and Russo, Frank A.},
  title   = {The benefit of hearing aids as measured by listening accuracy, subjective listening effort, and {fNIRS}},
  journal = {Trends in Hearing}, volume = {28}, year = {2024}
}

@article{winn2015spectral,
  author  = {Winn, Matthew B. and Edwards, Jan R. and Litovsky, Ruth Y.},
  title   = {The impact of auditory spectral resolution on listening effort revealed by pupil dilation},
  journal = {Ear and Hearing}, volume = {36}, number = {4}, pages = {e153--e165}, year = {2015}
}

@article{pichorafuller2016fuel,
  author  = {Pichora-Fuller, M. Kathleen and Kramer, Sophia E. and Eckert, Mark A. and others},
  title   = {Hearing impairment and cognitive energy: The framework for understanding effortful listening ({FUEL})},
  journal = {Ear and Hearing}, volume = {37}, pages = {5S--27S}, year = {2016}
}

@article{zekveld2018pupil,
  author  = {Zekveld, Adriana A. and Koelewijn, Thomas and Kramer, Sophia E.},
  title   = {The pupil dilation response to auditory stimuli: Current state of knowledge},
  journal = {Trends in Hearing}, volume = {22}, year = {2018}
}

@inproceedings{mittag2021nisqa,
  author    = {Mittag, Gabriel and Naderi, Babak and Chehadi, Assmaa and M{\"o}ller, Sebastian},
  title     = {{NISQA}: A deep {CNN}-self-attention model for multidimensional speech quality prediction with crowdsourced datasets},
  booktitle = {Proc. Interspeech}, pages = {2127--2131}, year = {2021}
}

@inproceedings{torcoli2024odaq,
  author    = {Torcoli, Matteo and Wu, Chih-Wei and Dick, Sascha and Williams, Phillip A. and Halimeh, Mhd Modar and Wolcott, William and Habets, Emanu{\"e}l A. P.},
  title     = {{ODAQ}: Open dataset of audio quality},
  booktitle = {Proc. IEEE ICASSP}, pages = {836--840}, year = {2024}
}

@article{hyvarinen2000ica,
  author  = {Hyv{\"a}rinen, Aapo and Oja, Erkki},
  title   = {Independent component analysis: Algorithms and applications},
  journal = {Neural Networks}, volume = {13}, number = {4--5}, pages = {411--430}, year = {2000}
}

@article{chen2010shrinkage,
  author  = {Chen, Yilun and Wiesel, Ami and Eldar, Yonina C. and Hero, Alfred O.},
  title   = {Shrinkage algorithms for {MMSE} covariance estimation},
  journal = {IEEE Trans. Signal Process.}, volume = {58}, number = {10}, pages = {5016--5029}, year = {2010}
}

@article{barachant2012riemannian,
  author  = {Barachant, Alexandre and Bonnet, St{\'e}phane and Congedo, Marco and Jutten, Christian},
  title   = {Multiclass brain--computer interface classification by {Riemannian} geometry},
  journal = {IEEE Trans. Biomed. Eng.}, volume = {59}, number = {4}, pages = {920--928}, year = {2012}
}

@article{vincent2006sdr,
  author  = {Vincent, Emmanuel and Gribonval, R{\'e}mi and F{\'e}votte, C{\'e}dric},
  title   = {Performance measurement in blind audio source separation},
  journal = {IEEE Trans. Audio, Speech, Language Process.}, volume = {14}, number = {4}, pages = {1462--1469}, year = {2006}
}

@article{delpy1988optical,
  author  = {Delpy, David T. and Cope, Mark and van der Zee, Pieter and Arridge, Simon and Wray, Susan and Wyatt, J.},
  title   = {Estimation of optical pathlength through tissue from direct time of flight measurement},
  journal = {Physics in Medicine \& Biology}, volume = {33}, number = {12}, pages = {1433}, year = {1988}
}

@article{scholkmann2010motion,
  author  = {Scholkmann, Felix and Spichtig, Stefan and Muehlemann, Thomas and Wolf, Martin},
  title   = {How to detect and reduce movement artifacts in near-infrared imaging using moving standard deviation and spline interpolation},
  journal = {Physiological Measurement}, volume = {31}, number = {5}, pages = {649--662}, year = {2010}

}

@article{scholkmann2014review,
  author  = {Scholkmann, Felix and Kleiser, Stefan and Metz, Andreas J. and Zimmermann, Raphael and Mata Pavia, Juan and Wolf, Ursula and Wolf, Martin},
  title   = {A review on continuous wave functional near-infrared spectroscopy and imaging instrumentation and methodology},
  journal = {NeuroImage}, volume = {85}, pages = {6--27}, year = {2014}
}

@article{simeone2025alpha,
  author  = {Simeone, Davide and Polo, Edoardo Maria and Barbieri, Riccardo and Paglialonga, Alessia},
  title   = {Dynamic assessment of listening effort by {EEG} alpha oscillations during an adaptive speech-in-noise test},
  journal = {Scientific Reports}, volume = {15}, year = {2025}
}

@article{antons2012eeg,
  author={Antons, Jan-Niklas and Schleicher, Robert and Arndt, Sebastian and M\"oller, Sebastian and Porbadnigk, Anne K. and Curio, Gabriel},
  title={Analyzing Speech Quality Perception Using Electroencephalography},
  journal={IEEE J. Sel. Topics Signal Process.}, volume={6}, number={6}, pages={721}, year={2012}}

@inproceedings{arndt2014interspeech,
  author    = {Sebastian Arndt and Markus Wenzel and Jan-Niklas Antons and Friedemann K{\"o}ster and Sebastian M{\"o}ller and Gabriel Curio},
title={A Next Step Towards Measuring Perceived Quality of Speech Through Physiology},
  booktitle={Proc. Interspeech}, pages={1998--2001}, year={2014}}

@article{miles2017coregistration,
  author={Miles, Kelly and McMahon, Catherine and Boisvert, Isabelle and Ibrahim, Ronny and de Lissa, Peter and Graham, Petra and Lyxell, Bj\"orn},
  title={Objective Assessment of Listening Effort: Coregistration of Pupillometry and {EEG}},
  journal={Trends in Hearing}, volume={21}, pages={1--13}, year={2017}}

@article{seifiala2020alpha,
  author={Seifi Ala, Tirdad and Graversen, Carina and Wendt, Dorothea and Alickovic, Emina and Whitmer, William M. and Lunner, Thomas},
  title={An Exploratory Study of {EEG} Alpha Oscillation and Pupil Dilation in Hearing-Aid Users During Effortful Listening to Continuous Speech},
  journal={PLOS ONE}, volume={15}, number={7}, pages={e0235782}, year={2020}}

@article{kilgour2018frechet,
  title={Fr\'echet Audio Distance: A Metric for Evaluating Music Enhancement Algorithms},
  author={Kilgour, Kevin and Zuluaga, Mauricio and Roblek, Dominik and Shari, Matthew},
  journal={CoRR},
  volume={abs/1812.08466},
  year={2018},
  archivePrefix={arXiv},
  eprint={1812.08466},
  url={https://arxiv.org/abs/1812.08466}
}

\end{document}